\documentclass[aps,prx,twocolumn,superscriptaddress,longbibliography]{revtex4-1}

\usepackage[english]{babel} 
\usepackage[latin1]{inputenc}
\usepackage{color}
\usepackage{graphicx}
\usepackage{amsmath,amssymb,amsfonts}
\usepackage[colorlinks=true,linkcolor=blue,urlcolor=blue,citecolor=blue]{hyperref}

\newcommand {\mc} {\mathcal}
\newcommand {\lan} {\left \langle}
\newcommand {\ran} {\right \rangle}

\DeclareMathOperator{\sign}{sign}

\begin{document}

\title{Minimal excitation states for heat transport in driven quantum Hall systems}

\author{Luca Vannucci}
\affiliation{Dipartimento di Fisica, Universit\`a di Genova, Via Dodecaneso 33, 16146, Genova, Italy}
\affiliation{CNR-SPIN, Via Dodecaneso 33, 16146, Genova, Italy}

\author{Flavio Ronetti}
\affiliation{Dipartimento di Fisica, Universit\`a di Genova, Via Dodecaneso 33, 16146, Genova, Italy}
\affiliation{CNR-SPIN, Via Dodecaneso 33, 16146, Genova, Italy}
\affiliation{Aix Marseille Univ, Universit\'e de Toulon, CNRS, CPT, Marseille, France}

\author{J\'er\^ome Rech}
\affiliation{Aix Marseille Univ, Universit\'e de Toulon, CNRS, CPT, Marseille, France}

\author{Dario Ferraro}
\affiliation{Aix Marseille Univ, Universit\'e de Toulon, CNRS, CPT, Marseille, France}

\author{Thibaut Jonckheere}
\affiliation{Aix Marseille Univ, Universit\'e de Toulon, CNRS, CPT, Marseille, France}

\author{Thierry Martin}
\affiliation{Aix Marseille Univ, Universit\'e de Toulon, CNRS, CPT, Marseille, France}

\author{Maura Sassetti}
\affiliation{Dipartimento di Fisica, Universit\`a di Genova, Via Dodecaneso 33, 16146, Genova, Italy}
\affiliation{CNR-SPIN, Via Dodecaneso 33, 16146, Genova, Italy}

\date{\today}

\begin{abstract}

We investigate minimal excitation states for heat transport into a fractional quantum Hall system driven out of equilibrium by means of time-periodic voltage pulses. A quantum point contact allows for tunneling of fractional quasi-particles between opposite edge states, thus acting as a beam splitter in the framework of the electron quantum optics. Excitations are then studied through heat and mixed noise generated by the random partitioning at the barrier. It is shown that levitons, the single-particle excitations of a filled Fermi sea recently observed in experiments, represent the cleanest states for heat transport, since excess heat and mixed shot noise both vanish \textit{only} when Lorentzian voltage pulses carrying integer electric charge are applied to the conductor. This happens in the integer quantum Hall regime and for Laughlin fractional states as well, with no influence of fractional physics on the conditions for clean energy pulses.
In addition, we demonstrate the robustness of such excitations to the overlap of Lorentzian wavepackets. Even though mixed and heat noise have nonlinear dependence on the voltage bias, and despite the non-integer power-law behavior arising from the fractional quantum Hall physics, an arbitrary superposition of levitons always generates minimal excitation states.

\end{abstract}

\maketitle

\section{Introduction}
\label{sec:intro}

The emerging field of electron quantum optics aims at manipulating electrons one by one in ballistic, coherent conductors \cite{Bocquillon14}. In this way it is possible to reproduce quantum-optical experiments and setups in solid state devices, using fermionic degrees of freedom (electrons in mesoscopic systems) instead of bosonic ones (photons in waveguides and optical cavities).
For this purpose a huge effort was committed towards the realization of single-electron sources, which clearly represent a crucial building block to perform any quantum-optical experiment in electronic systems.
The first proposal to extract a single electron out of the filled Fermi sea was theoretically discussed by B\"uttiker and collaborators and is known as the mesoscopic capacitor \cite{buttiker93_PLA,buttiker93_JPCM}. It consists of a quantum dot connected to a two-dimensional electron gas through a quantum point contact (QPC), where a periodic drive of the energy levels of the dot leads to the alternate injection of an electron and a hole into the system for each period of the drive \cite{Gabelli06,Feve07}.
An equally effective yet conceptually simpler idea to conceive single-electron excitations was discussed by Levitov and coworkers, who showed how to excite a single electron above the Fermi sea applying well defined voltage pulses to a quantum conductor \cite{levitov96,ivanov97,keeling06}. While a generic voltage drive would generate an enormous amount of particle-hole pairs in the conductor, a Lorentzian drive carrying an integer amount of electrons per period produces particle-like excitations only (such single-electron excitations are now dubbed levitons). Although challenging, the idea of voltage pulse generation proved to be simpler than the mesoscopic capacitor as it does not involve delicate nanolithography, thus drastically simplifying the fabrication process of the single-electron gun.
Hanbury-Brown and Twiss partitioning experiments and Hong-Ou-Mandel interferometers with single-electron sources were experimentally reported using both the mesoscopic capacitor and levitons \cite{Bocquillon12,Bocquillon13,dubois13-nature}.
Several proposal have been formulated to use levitons as flying qubits for the realization of quantum logic gates \cite{Glattli16_pss}, as a source of entanglement in Mach-Zehnder interferometers \cite{Dasenbrook15,Dasenbrook16_NJP,Dasenbrook16_PRL}, or to conceive zero-energy excitation carrying half the electron charge \cite{Moskalets16_PRL}.
Quantum tomography protocols for electron states were also theoretically developed \cite{Grenier11,Ferraro13} and implemented using levitons as a benchmark quantum state \cite{Jullien14}.
Moreover, in a recent work it was shown that conditions for minimal excitations are unaffected in the fractional quantum Hall (FQH) regime \cite{Rech16}. Here the notion of leviton was extended to interacting systems of the Laughlin sequence, and it was demonstrated that Lorentzian pulses carrying \textit{integer} charge represent the cleanest voltage drive despite the fundamental carriers being quasi-particles with fractional charge and statistics \cite{Laughlin83,Saminadayar97,dePicciotto97}.

Despite several challenging and fascinating problems concerning charge transport properties, electric charge is far from being the only interesting degree of freedom we should look at in the framework of electron quantum optics. Energy, for instance, can be coherently transmitted over very long distances along the edge of quantum Hall systems, as was experimentally proved by Granger \textit{et al.} \cite{Granger09}.
This observation is of particular interest, as typical dimensions of chips and transistors are rapidly getting smaller and smaller due to the great technological advance during the last decades. Indeed, the problem of heat conduction and manipulation at the nanoscale has become more actual than ever \cite{Giazotto06}, as demonstrated by great recent progress in the field of quantum thermodynamics. 
Topics like quantum fluctuation-dissipation theorems \cite{Campisi09,Campisi11,Averin10,Whitney16,moskalets14}, energy exchanges in open quantum systems \cite{Carrega15,Carrega16}, energy dynamics and pumping at the quantum level \cite{Ludovico14,Ludovico16,Calzona16,Calzona17,Ronetti17}, coherent caloritronics \cite{Giazotto12,Fornieri16}, and thermoelectric phenomena \cite{Benenti16,SanchezD16,Whitney14} have all been extensively investigated, in an attempt to extend the known concepts of thermodynamics to the quantum realm.
In this context, a particular emphasis has been focused on the role of quantum Hall edge states both from the theoretical \cite{Grosfeld09,Arrachea11,Aita13,SanchezR15_PRL,vannucci15,Samuelsson16,SanchezR15_PRL} and experimental point of view \cite{Granger09,Altimiras10,Altimiras12,Venkatachalam12,Gurman12,Banerjee16_arxiv}.

A natural question immediately arises when one considers energy dynamics in electron quantum optics, namely what kind of voltage drive gives rise to minimal excitation states for heat transport in mesoscopic conductors. This is the fundamental question we try to answer in this paper. To this end, we study heat conduction along the topologically-protected chiral edge states of the quantum Hall effect. We analyze heat current fluctuations as well as mixed charge-heat correlations \cite{Crepieux14,Crepieux16} when periodic voltage pulses are sent to the conductor and partitioned off a QPC \cite{dubois13-nature}.
Starting from the DC regime of the voltage drive, where simple relations between noises and currents can be derived in the spirit of the celebrated Schottky's formula \cite{Schottky18,blanter00}, we introduce the \textit{excess signals} for charge, heat and mixed fluctuations, which basically measure the difference between the zero-frequency noises in an AC-driven system and their respective reference signals in the DC configuration. The vanishing of excess heat and mixed noise is thus used to flag the occurrence of a minimal excitation state for heat transport in the quantum Hall regime. 
With this powerful tool we demonstrate that minimal noise states for heat transport can be achieved only when the voltage drive takes the form of Lorentzian pulses carrying an integer multiple of the electron charge, i.e.~when levitons are injected into the quantum Hall edge states.
We study this problem both in the integer regime and in the FQH regime, where strong interactions give rise to the fractional properties of quasi-particle excitations. Our results show a striking robustness against interactions, since integer levitons still represent minimal excitation states despite the highly non-linear physics occurring at the QPC due to the peculiar collective excitations of the FQH state.

Having recognized levitons as the fundamental building block for heat transport, we then turn to the second central issue of this paper, which deals with the robustness of multiple overlapping Lorentzian pulses as minimal excitation states. Indeed, Levitov and collaborators demonstrated that $N$ levitons traveling through a quantum conductor with transmission $\mc T<1$ represent $N$ independent attempts to pass the barrier, with the total noise not affected by the overlap between their wavepackets. This is no more guaranteed when we look for quantities which, unlike the charge current and noise, have a non-linear dependence on the voltage bias.
Two types of nonlinearities are considered in this work. The first one comes from the mixed and heat shot noise, whose behaviors are $\sim V^2(t)$ and $\sim V^3(t)$ respectively in Fermi liquid systems. The second one is a natural consequence of FQH physics, which give rise to exotic power laws with non-integer exponents.
We show that, while currents and noises are sensitive to the actual number of particles sent to the QPC, excess signals always vanish for arbitrary superposition of integer levitons. One then concludes that levitons show a remarkable stability even with regard to heat transport properties, combined with the equally surprising robustness in the strongly-correlated FQH liquid. This provides further evidence of the uniqueness of the leviton state in the quantum Hall regime.

The content of the paper is organized in the following way. The model for the FQH bar in presence of a periodic voltage drive is presented in Sec.~\ref{sec:model}, followed by the evaluation of expectation values for noises and currents in Sec.~\ref{sec:current_noise}. Excess signals are then introduced in Sec.~\ref{sec:excess_signal}, where results concerning their vanishing for quantized lorentizan voltage pulses are also presented. Finally we analyze the problem of multiple levitons in Sec.~\ref{sec:multiple}, before drawing our conclusions in Sec.~\ref{sec:conclusions}. Three Appendices are devoted to the technical details of the calculations.

\section{Model}
\label{sec:model}

We consider a quantum Hall system with filling factor $\nu=1/(2n+1)$, $n\in \mathbb {N}$. The special case $n=0$ corresponds to the integer quantum Hall regime at $\nu=1$, where the single chiral state on each edge is well described by a one-dimensional Fermi liquid theory. Conversely, values $n>0$ describe a fractional system in the Laughlin sequence \cite{Laughlin83}, with still one chiral mode per edge. The free Hamiltonian modeling right-moving and left-moving states on opposite edges is $H_0=H_R+H_L$ with \cite{Wen95}
\begin{equation}
	\label{eq:H_R/L}
	\quad H_{R/L} = \frac{v}{4\pi} \int dx \left[ \partial_x \phi_{R/L}(x) \right]^2 ,
\end{equation}
where $\phi_{R/L}(x)$ are bosonic fields satisfying $[\phi_{R/L}(x), \phi_{R/L}(y)] = \pm i \pi \sign(x-y)$. In Eq.~\eqref{eq:H_R/L} and throughout the rest of the paper we set $\hbar=1$. The parameter $v$ is the propagation velocity for the chiral edge states, meaning that free bosonic fields evolve in time as $\phi_{R/L}(x,t) = \phi_{R/L}(x \mp vt,0)$.
One can relate the bosonic description to creation and annihilation of quasi-particles through bosonization identities \cite{miranda03,vondelft}
\begin{equation}
	\label{eq:bosonization}
	\psi_{R/L} (x) = \frac{f_{R/L}}{\sqrt{2\pi a}} e^{\pm i k_{\rm F} x} e^{-i \sqrt \nu \phi_{R/L}(x)} ,
\end{equation}
where the field $\psi$ represents annihilation of a quasi-particle with fractional charge $-e^*=-\nu e$ ($e>0$). The parameter $a$ in Eq.~\eqref{eq:bosonization} is a short distance cutoff and $f_{R/L}$ are the so called Klein factors. They will be omitted in the rest of the paper, as they do not affect our calculations.
In the bosonic formalism, quasi-particle density operators are given by
\begin{equation}
	\rho_{R/L} (x) = \mp \frac{\sqrt{\nu}}{2\pi} \partial_x \phi_{R/L}(x) .
\end{equation}

\begin{figure}
	\centering
	\includegraphics[width=0.9\linewidth]{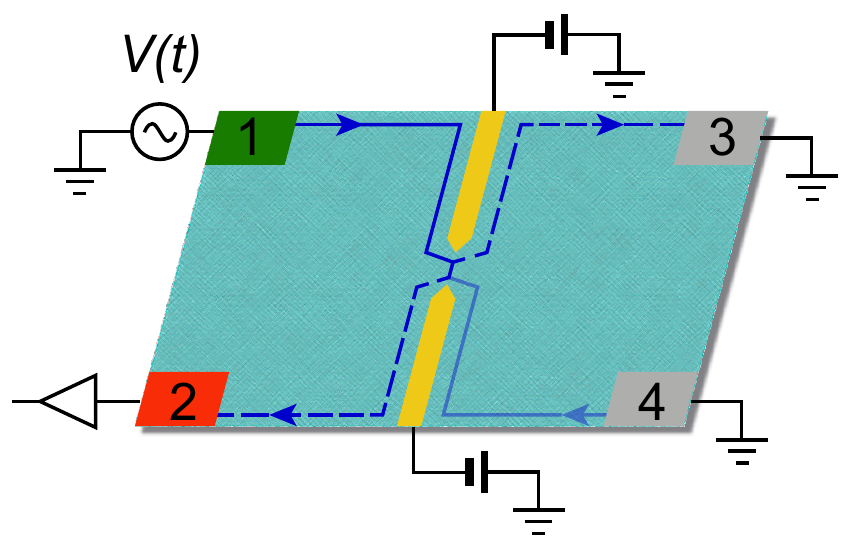}
	\caption{(Color online) Fractional quantum Hall liquid in a four-terminal setup. Two gate voltages create a quantum point contact at $x=0$, allowing for tunneling between opposite edges. A periodic bias $V(t)$ is applied to contact 1, while contacts 3 and 4 are grounded. Backscattered currents and their fluctuations are detected in contact 2.}
	\label{fig:setup}
\end{figure}

Quasi-particle tunneling occurs at $x=0$ due to the presence of a QPC, schematically depicted in Fig.~\ref{fig:setup}. This is modeled through the tunneling Hamiltonian 
\begin{equation}
	\label{eq:h_tun}
	H_{\rm t} = \Lambda e^{i e^* \int_{0}^t V(t') dt'}  \psi^\dag_R(0,t) \psi_L(0,t) + \mathrm{h.c.} ,
\end{equation}
with $\Lambda$ the constant tunneling strength. Here the phase $e^{i e^* \int_{0}^t V(t') dt'}$ takes into account the presence of a periodic voltage bias $V(t) = V_{\rm DC} + V_{\rm AC}(t)$ in terminal 1 (see Fig.~\ref{fig:setup}), where $V_{\rm DC}$ is a time-independent DC component and $V_{\rm AC}(t)$ is a pure periodic AC signal, i.e.~$\int_0^T V_{\rm AC}(t') dt'=0$ with $T=2\pi/\omega$ the period of the drive. This results in a phase shift of $\psi_{R}(0,t)$, as can be inferred by solving the equation of motion for the bosonic field subjected to an additional voltage drive (see Appendix \ref{app:eq_motion}).
The periodic phase $e^{-i \varphi(t)}$, with $\varphi(t) = e^* \int_{0}^t V_{\rm AC}(t') dt'$, will be conveniently handled through the Fourier series $e^{-i \varphi(t)} = \sum_{l=-\infty}^{+\infty} p_l e^{-i l \omega t}$, with coefficients $p_l$ given by
\begin{equation}
	\label{eq:Fourier_coeff}
	p_l =  \int_{0}^{T} \frac{dt}{T} e^{i l \omega t} e^{-i \varphi(t)} .
\end{equation}
Each coefficient $p_l$ represents the probability amplitude for an electron to emit or absorb $l$ energy quanta from the electromagnetic field \cite{dubois13-nature}. Details of the calculation of Eq.~\eqref{eq:Fourier_coeff} for the voltage drives considered in this work are given in Appendix \ref{app:fourier}.

In the following we will discuss charge and heat current fluctuations as a function of the total charge $q$ injected during one period of the drive, in units of the elementary charge $e$. For a FQH edge state with conductance $G=\nu \frac{e^2}{2\pi}$ this reads
\begin{equation}
	q = \frac 1 e \int_0^T G V(t') dt' = \frac{e^*}{\omega} V_{\rm DC} .
\end{equation}

\section{Zero-frequency heat and mixed noise}
\label{sec:current_noise}

Operators for charge and heat currents backscattered off the barrier and detected in terminal 2 are defined as \cite{Callen}
\begin{subequations}
\label{eq:def_J}
\begin{align}
	J_C(t) & = e^* \dot N_L(t) , \\
	J_Q(t) & = \dot H_L(t) - \mu \dot N_L(t) ,
\end{align}
\end{subequations}
where $N_L(t) = \int dx \rho_L(x,t)$ is the number of quasi-particles in the left-moving edge, $\mu=v k_{\rm F}$ is the chemical potential in contact 2 and $\dot H_L$ follows from Eq.~\eqref{eq:H_R/L}. Our focus will be on the zero-frequency component of the power spectra
\begin{equation}
	\label{eq:def_S}
	\mc S_{ij} = 2 \int_0^T \frac{dt}{T} \int_{-\infty}^{+\infty} dt' \lan \Delta J_i(t) \Delta J_j(t') \ran ,
\end{equation}
with $i,j=\{C,Q\}$ and the operator $\Delta J_i(t) = J_i(t) - \lan J_i(t) \ran$ describing charge and heat current fluctuations. We will use the short-hand notation $\mc S_C = S_{CC}$ for the charge shot noise, $\mc S_X = S_{CQ}$ for the mixed correlator and $\mc S_Q = S_{QQ}$ for the heat noise. 

We resort to the Keldysh non-equilibrium formalism \cite{Rammer,Martin_Houches} for the calculation of expectation values, whose details are reported in Appendix \ref{app:calculations}. To lowest order in the tunneling we obtain
\begin{align}
	\mc S_C	& = 4 (e^*)^2 |\lambda|^2 \int_0^T \frac{dt}{T} \int_{-\infty}^{+\infty} dt' \cos \left[e^* \int_{t'}^t d t'' V(t'') \right] \nonumber \\
	\label{eq:S_C(t)}
	& \quad \times e^{2\nu \mc G(t'-t)} , \\
	\mc S_X & = 4 e^* |\lambda|^2 \int_0^T \frac{dt}{T} \int_{-\infty}^{+\infty} dt' \sin \left[e^* \int_{t'}^t d t'' V(t'') \right] \nonumber \\
	\label{eq:S_X(t)}
	& \quad \times e^{\nu \mc G(t'-t)} \partial_{t'} e^{\nu \mc G(t'-t)}  ,\\
	\mc S_Q	& = 4 |\lambda|^2 \int_0^T \frac{dt}{T} \int_{-\infty}^{+\infty} dt' \cos \left[e^* \int_{t'}^t d t'' V(t'') \right] \nonumber \\
	\label{eq:S_Q(t)}
	& \quad \times e^{\nu \mc G(t'-t)} \partial_t \partial_{t'} e^{\nu \mc G(t'-t)} .
\end{align}
with $\mc G(\tau) = \lan \left[ \phi_{R/L}(0,\tau) - \phi_{R/L}(0,0)\right] \phi_{R/L}(0,0) \ran$ the bosonic correlation function, equal for both right-moving and left-moving modes, and $\lambda=\Lambda/(2\pi a)$ the reduced tunneling constant. Introducing the Fourier transform of $e^{g \mc G(\tau)}$, namely $\hat P_g (E) = \int d\tau e^{i E \tau} e^{g \mc G(\tau)}$, and the series representation for $e^{-i \varphi(t)}$ we get 
\begin{align}
	\label{eq:S_C_finite_temp}
	\mc S_C & = 2 (e^*)^2 |\lambda|^2 \sum_{l} |p_l|^2 \nonumber \\
	& \quad \times \left\{ \hat P_{2\nu} \left[ (q+l)\omega \right] + \hat P_{2\nu} \left[ -(q+l)\omega \right] \right\} , \\
	\label{eq:S_X_finite_temp}
	\mc S_X & = e^* \omega |\lambda|^2 \sum_{l} |p_l|^2 (q+l) \nonumber \\
	& \quad \times \left\{ \hat P_{2\nu} \left[ (q+l)\omega \right] + \hat P_{2\nu} \left[ -(q+l)\omega \right] \right\} , \\
	\label{eq:S_Q_finite_temp}
	\mc S_Q & = |\lambda|^2 \sum_{l} \left|p_l\right|^2 \left[ \frac{2\pi^2 \nu^2}{1+2\nu} \theta^2 + \frac{1+\nu}{1+2\nu} (q+l)^2 \omega^2 \right]\nonumber \\
	& \quad \times \left\{ \hat P_{2\nu} \left[ (q+l)\omega \right] + \hat P_{2\nu} \left[ -(q+l)\omega \right] \right\} .
\end{align}
for the zero-frequency component of the noises. In particular, at temperature $\theta=0$ one has
\begin{equation}
	\label{eq:Pg_zero_temp}
	\hat P_g(E) = \frac{2\pi}{\Gamma(g) \omega_{\rm c}} \left| \frac{E}{\omega_{\rm c}}\right|^{g-1} \Theta(E) ,
\end{equation}
with $\omega_{\rm c}=v/a$ the high energy cutoff and $\Theta(E)$ the Heaviside step function. The noises then reduce to
\begin{align}
	\label{eq:S_C}
	\mc S_C & = \frac{(e^*)^2}{\omega} |\lambda|^2 \frac{4\pi}{\Gamma(2\nu)} \left(\frac{\omega}{\omega_{\rm c}}\right)^{2\nu} \sum_{l} |p_l|^2 |q+l|^{2\nu-1} , \\
	\label{eq:S_X}
	\mc S_X & = e^* |\lambda|^2 \frac{2\pi}{\Gamma(2\nu)} \left(\frac{\omega}{\omega_{\rm c}}\right)^{2\nu} \sum_{l} |p_l|^2 |q+l|^{2\nu} \sign(q+l) , \\
	\label{eq:S_Q}
	\mc S_Q & = \omega |\lambda|^2 \frac{2\pi (1+\nu)}{\Gamma(2\nu) (1+2\nu)} \left(\frac{\omega}{\omega_{\rm c}}\right)^{2\nu} \sum_{l} |p_l|^2 |q+l|^{2\nu+1} . 
\end{align}
Equation \eqref{eq:Pg_zero_temp} and subsequent Eqs.~\eqref{eq:S_C}, \eqref{eq:S_X} and \eqref{eq:S_Q} show the familiar power-law behavior of the Luttinger liquid \cite{Fisher97,giamarchi}.

It is instructive to calculate also the averaged charge and heat currents flowing into contact 2 for later use. In this case Keldysh formalism yields
\begin{align}
	\label{eq:J_C(t)}
	\lan J_C(t) \ran
	& = 2i e^* |\lambda|^2 \int_{0}^{+\infty} d\tau \sin \left[ e^* \int_{t-\tau}^t dt'' V(t'') \right] \nonumber \\
	& \quad \times \left[ e^{ 2 \nu \mc G(\tau)} - e^{ 2 \nu \mc G(-\tau)} \right] ,\\
	\label{eq:J_Q(t)}
	\lan J_Q(t) \ran
	& = i |\lambda|^2 \int_{0}^{+\infty} d\tau \cos \left[ e^* \int_{t-\tau}^t dt'' V(t'') \right] \nonumber \\
	& \quad \times \left[ \partial_\tau e^{ 2 \nu \mc G(\tau)} - \partial_\tau e^{ 2 \nu \mc G(-\tau)} \right] .
\end{align}
The DC component of charge and heat currents in presence of the periodic drive is then obtained by averaging over one period $T$. Below we give the results for the zero-temperature signals, using the symbol $\overline{\lan \dots \ran}$ to denote the time average $\int_0^T \frac{dt}{T} \lan \dots \ran$:
\begin{align}
	\overline{\lan J_C(t) \ran} & = \frac{|\lambda|^2 e^*}{\omega} \frac{2\pi}{\Gamma(2\nu)} \left( \frac{\omega}{\omega_{\rm c}}\right)^{2\nu} \sum_{l} |p_l|^2 |q+l|^{2\nu-1} \nonumber \\
	\label{eq:avg_J_C}
	& \quad \times \sign(q+l) , \\
	\label{eq:avg_J_Q}
	\overline{\lan J_Q(t) \ran} & = |\lambda|^2 \frac{\pi}{\Gamma(2\nu)} \left( \frac{\omega}{\omega_{\rm c}}\right)^{2\nu} \sum_{l} |p_l|^2 |q+l|^{2\nu} .
\end{align}
Intermediate steps of the calculation and finite-temperature expressions for $\overline{\lan J_C(t) \ran}$ and $\overline{\lan J_Q(t) \ran}$ are developed in Appendix \ref{app:calculations}.

\section{Excess signals and noiseless drive}
\label{sec:excess_signal}

\subsection{From Schottky formula to the AC regime}

We start the discussion considering a DC-biased conductor, i.e.~$V(t)=V_{\rm DC}$ with $V_{\rm AC}(t)=0$. Such a situation entails that Fourier coefficients in Eq.~\eqref{eq:Fourier_coeff} reduce to $p_l = \delta_{l,0}$. In this case, charge current and noise at temperature $\theta=0$ are linked by \cite{Kane94,Saminadayar97,dePicciotto97}
\begin{align}
	\label{eq:schottky_C}
	\mc S_C & = 2e^* \lan J_C \ran ,
\end{align}
which can be easily checked from our formulas. Equation \eqref{eq:schottky_C} is a manifestation of the Schottky relation for a system with fractionally charged carriers \cite{Schottky18,blanter00}. It is linked to the fact that transmission of uncorrelated single-particle excitations through a barrier is described by Poisson distribution, hence the proportionality between shot noise and charge current. Interestingly, similar expressions can be derived relating mixed and heat noise to the heat current for a DC bias. From Eq.~\eqref{eq:avg_J_Q} and assuming $V_{\rm DC}>0$, one gets the following formula for the heat current
\begin{equation}
	\label{eq:DC_heat_current}
	\lan J_Q \ran=\left|\lambda\right|^2\frac{\pi}{\Gamma{\left(2\nu\right)}}\left(\frac{e^* V_{\rm DC}}{\omega_c}\right)^{2\nu} .
\end{equation} 
Similarly, mixed and heat noise are obtained from Eqs.~\eqref{eq:S_X} and \eqref{eq:S_Q} with the condition $p_l = \delta_{l,0}$. They read
\begin{align}
	\label{eq:DC_mixed}
	\mc S_X &= e^{*}\left|\lambda\right|^2\frac{2\pi}{\Gamma{\left(2\nu\right)}}\left(\frac{e^* V_{\rm DC}}{\omega_c}\right)^{2\nu}, \\
	\label{eq:DC_heat}
	\mc S_Q &= e^{*}V_{DC}\left|\lambda\right|^2\frac{2\pi\left(1+\nu\right)}{\Gamma{\left(2\nu\right)}\left(1+2\nu\right)}\left(\frac{e^* V_{\rm DC}}{\omega_c}\right)^{2\nu} .
\end{align}
Comparing the last three results we immediately notice a proportionality between $\mc S_X$, $\mc S_Q$ and $\lan J_Q \ran$, namely
\begin{align}
	\label{eq:schottky_X}
	\mc S_X & = 2e^* \lan J_Q \ran , \\
	\label{eq:schottky_Q}
	\mc S_Q & = 2e^* \frac{1+\nu}{1+2\nu} V_{DC} \lan J_Q \ran .
\end{align}
Equations \eqref{eq:schottky_X} and \eqref{eq:schottky_Q} are generalizations of Schottky's formula to the heat and mixed noise.
They show that the uncorrelated backscattering of Laughlin quasi-particles at the QPC leaves Poissonian signature in heat transport properties also, in addition to the well-known Poissonian behavior of the charge shot noise described by Eq.~\eqref{eq:schottky_C}. This holds both in a chiral Fermi liquid (i.e.~at $\nu=1$, when tunneling involves integer electrons only) and in the FQH regime, with proportionality constants governed by the filling factor $\nu$.
Similar relations for transport across a quantum dot were recently reported \cite{Crepieux14,eymeoud16}.

In general, the Schottky relation breaks down in the AC regime, since the oscillating drive excites particle-hole pairs contributing to transport.
Nevertheless, when a single electron is extracted from the filled Fermi sea we expect the photon-assisted zero-frequency shot noise to match the lower bound set by Schottky's Poissonian DC relation. Thus the quantity
\begin{equation}
	\label{eq:def_exc_S_C}
	\Delta \mc S_C = \mc S_C - 2e^* \overline{\lan J_C(t) \ran} ,
\end{equation}
which we call \textit{excess charge noise}, vanishes in the presence of a minimal excitation state as already mentioned in earlier works \cite{dubois13,dubois13-nature,Rech16}. For completeness, we quote its expression at zero temperature:
\begin{equation}
	\label{eq:exc_S_C}
	\Delta \mc S_C = \frac{(e^*)^2}{\omega} |\lambda|^2 \frac{8\pi}{\Gamma(2\nu)} \left(\frac{\omega}{\omega_{\rm c}}\right)^{2\nu} \sum_{l<-q} |p_l|^2 |q+l|^{2\nu-1} .
\end{equation}

We now address the central quantities of interest for the present paper. Equation \eqref{eq:schottky_X}, representing a proportionality between the mixed charge-heat correlator $\mc S_X$ and the heat current for a DC voltage drive governed by the charge $e^*$, leads us to introduce the \textit {excess mixed noise} given by
\begin{equation}
	\label{eq:def_exc_S_X}
	\Delta \mc S_X = \mc S_X - 2e^* \overline{\lan J_Q(t)\ran} .
\end{equation}
As for $\Delta \mc S_C$, this quantity measures the difference between the noise in presence of a generic periodic voltage drive and the DC reference value. Using the results of Sec.~\ref{sec:current_noise} the excess mixed noise reads
\begin{equation}
	\label{eq:exc_S_X}
	\Delta \mc S_X = -e^* |\lambda|^2 \frac{4\pi}{\Gamma(2\nu)} \left(\frac{\omega}{\omega_{\rm c}}\right)^{2\nu} \sum_{l<-q} |p_l|^2 |q+l|^{2\nu} .
\end{equation}
The vanishing of $\Delta \mc S_X$ should highlight an energetically clean pulse, for which the mixed noise reaches the minimal value $\mc S_X = 2e^* \overline{\lan J_Q(t)\ran}$ expected from Schottky's formula for the mixed noise Eq.~\eqref{eq:schottky_X}.
With a very similar procedure it is possible to extract the excess component of the zero-frequency heat noise due to the time dependent drive. Equation \eqref{eq:schottky_Q} states that $\mc S_Q$ is proportional to the heat current multiplied by the voltage bias in the DC limit. In view of this consideration we define the \textit{excess heat noise}
\begin{equation}
	\label{eq:def_exc_S_Q}
	\Delta \mc S_Q = \mc S_Q - 2e^* \frac{1+\nu}{1+2\nu} \overline{V(t) \lan J_Q(t) \ran} .
\end{equation}
The time-averaged value of $V(t) \lan J_Q(t) \ran$ can be calculated from Eq.~\eqref{eq:J_Q(t)} using the relation $e^* V(t) e^{-i\varphi(t)} = (\omega q + i \partial_t) e^{-i\varphi(t)}$. Then from the above definition we get
\begin{equation}
	\label{eq:exc_S_Q}
	\Delta \mc S_Q = \omega |\lambda|^2 \frac{4\pi(1+\nu)}{\Gamma(2\nu)(1+2\nu)} \left(\frac{\omega}{\omega_{\rm c}}\right)^{2\nu} \sum_{l<-q} |p_l|^2 |q+l|^{2\nu+1} .
\end{equation}

\subsection{Physical content of the excess signals}

Let us now look for the physics described by Eqs.~\eqref{eq:exc_S_X} and \eqref{eq:exc_S_Q}. Once again, it is enlightening to start from the analogy with the charge shot noise. In the $\nu=1$ quantum Hall state, described by a one-dimensional chiral Fermi liquid, the excess charge noise $\Delta \mc S_C$ is proportional to the number of holes $N_h$ induced in the Fermi sea by the voltage drive. One has
\begin{align}
	\label{eq:N_h}
	N_h(t) & = \sum_{k=-\infty}^{+\infty} n_{\rm F}(vk) \lan c_k(t) c^\dag_k(t) \ran \nonumber \\
	& = \frac{v^2}{(2\pi a)^2} \int d\tau' \int d\tau \, e^{ - i e \int_{\tau'-\tau}^{\tau'} dt' V(t') } e^{2 \mc G(\tau)} \nonumber \\
	& \propto \mc S_C - 2e \overline{\lan J_C(t)\ran} ,
\end{align}
where $n_{\rm F}(E) = \Theta(-E)$ is the Fermi distribution at zero temperature.
A similar relation is obtained in the fractional regime when we introduce the effective tunneling density of states $\mc D_\nu(E)$ of the chiral Luttinger liquid, which is reported in Appendix \ref{app:calculations}. The number of quasi-holes in the FQH liquid reads
\begin{align}
	\label{eq:N_qh}
	N_{qh} & = \sum_{k=-\infty}^{+\infty} \frac{\omega_{\rm c}}{2\pi} \mc D_\nu(vk) n_{\rm F}(vk) \lan c_{qp,k}(t) c^\dag_{qp,k}(t) \ran \nonumber \\
	& = \frac{v^2}{(2\pi a)^2} \int d\tau' \int d\tau e^{ - i e \int_{\tau'-\tau}^{\tau'} dt' V(t') } e^{2 \nu \mc G(\tau)} \nonumber \\
	& \propto \mc S_C - 2e^* \overline{\lan J_C(t)\ran} .
\end{align}
It is worth noticing that Eqs.~\eqref{eq:N_h} and \eqref{eq:N_qh} hold in an unperturbed system without tunneling between opposite edges. The shot-noise induced by the presence of the QPC can thus be viewed as a probe for the number of holes (or quasi-holes in the case of a fractional filling) generated by the AC pulses.

We now consider the energy associated with hole-like excitations for a generic filling factor of the Laughlin sequence, that reads
\begin{equation}
	E_{qh} = - \sum_{k=-\infty}^{+\infty} \frac{\omega_{\rm c}}{2\pi} \mc D_\nu(vk) n_{\rm F}(vk) vk \lan c_{qp,k}(t) c^\dag_{qp,k}(t) \ran .
\end{equation}
This quantity can be written as
\begin{align}
	E_{qh}
	& = \frac{i}{2} \frac{v^2}{(2\pi a)^2} \int d\tau' \int d\tau \, e^{- i e^* \int_{\tau'-\tau}^{\tau'} dt' V(t')} \partial_\tau e^{2\nu \mc G(\tau)} \nonumber \\
	& = \frac{1}{2} \frac{v^2}{(2\pi a)^2} \int d\tau' \int d\tau \, \left\{\sin \left[ e^* \int_{\tau'-\tau}^{\tau'} dt' V(t')\right] \right. \nonumber \\
	& \quad \left. + i \cos \left[ e^* \int_{\tau'-\tau}^{\tau'} dt' V(t')\right] \right\} \partial_\tau e^{2\nu \mc G(\tau)} .
\end{align}
Then, comparing this result with Eqs.~\eqref{eq:S_X(t)} and \eqref{eq:J_Q(t)} we find that $\Delta \mc S_X$ measures the energy associated with the unwanted quasi-holes generated through the periodic voltage drive, namely
\begin{equation}
	\label{eq:E_qh}
	E_{qh} \propto - \mc S_X + 2e^* \overline{\lan J_Q(t)\ran} = - \Delta \mc S_X .
\end{equation}
This accounts for the negative value of $\Delta \mc S_X$ arising from Eq.~\eqref{eq:exc_S_X}.
A similar relation involving the sum of the squared energy for each value of $k$ holds for $\Delta \mc S_Q$:
\begin{align}
	\label{eq:E2_qh}
	\sum_{k=-\infty}^{+\infty} \frac{\omega_{\rm c}}{2\pi} \mc D_\nu(vk) n_{\rm F}(vk) (vk)^2 \lan c_{qp,k}(t) c^\dag_{qp,k}(t) \ran \propto \Delta \mc S_Q  .
\end{align}

\begin{figure}
	\centering
	\includegraphics[width=\linewidth]{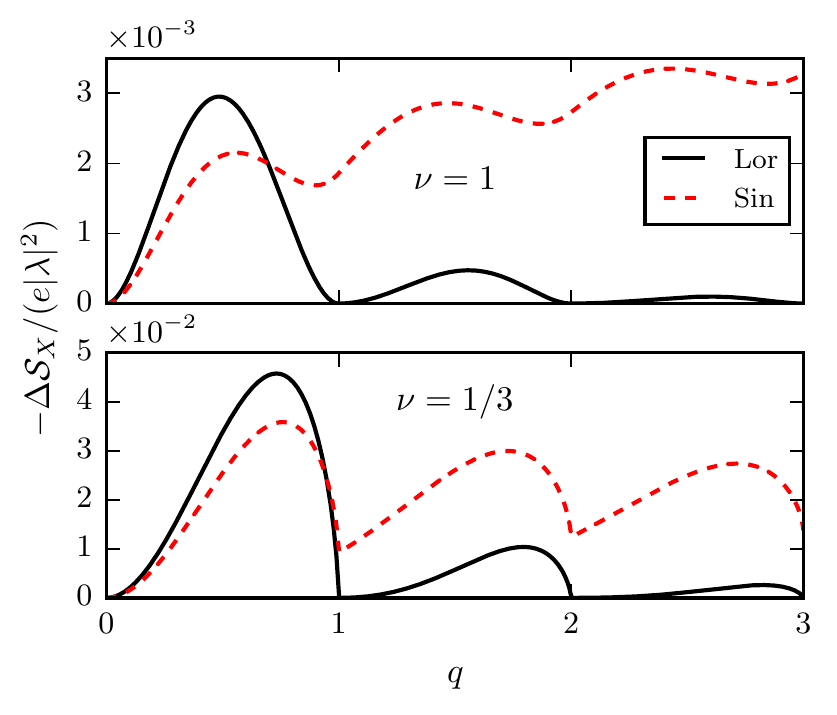}
	\caption{(Color online) Excess mixed noise $-\Delta \mc S_X$ as a function of the charge per period $q$ at zero temperature. The high energy cutoff is set to $\omega_{\rm c} = 10 \, \omega$. Behavior for Lorentzian pulses (full black line) and sinusoidal voltage drive (dashed red line) is reported.}
	\label{fig:Sx}
\end{figure}

In Fig.~\ref{fig:Sx} we show the behavior of the excess mixed noise as a function of the charge $q$ injected during one period $T$. Notice that we normalize $\Delta \mc S_X$ by a negative quantity, in order to deal with a positive function. Two types of bias are considered: a sinusoidal drive and a train of Lorentzian pulses given respectively by
\begin{align}
	V_{\rm sin}(t) & = V_{\rm DC} [1-\cos(\omega t)] ,\\
	V_{\rm Lor}(t) & = \frac{V_{\rm DC}}{\pi} \sum_{k} \frac{\eta}{\eta^2 + (t/T-k)^2} , 
\end{align}
with $\eta=W/T$ the ratio between the half width $W$ at half maximum of the Lorentzian peak and the period $T$. The former is representative of all kinds of non-optimal voltage drive, while the latter is known to give rise to minimal charge noise both at integer \cite{keeling06} and fractional \cite{Rech16} fillings. We will set $\eta=0.1$, a value lying in the range investigated by experiments \cite{dubois13-nature}. At $\nu=1$, both curves display local minima whenever $q$ assumes integer values. However, while the sinusoidal drive always generates an additional noise with respect to the reference Schottky value $2e^* \overline{\lan J_Q(t)\ran}$, the Lorentzian signal drops to zero for $q \in \mathbb N$, indicating that the mixed noise $\mc S_X$ due to levitons exactly matches the Poissonian value set by Eq.~\eqref{eq:schottky_X}. Since the excess mixed noise is linked to the unwanted energy introduced into the system as a result of hole injection [see Eq.~\eqref{eq:E_qh}], Fig.~\ref{fig:Sx} shows that there is no hole-like excitation carrying energy in our system.
The bottom panel of Fig.~\ref{fig:Sx} shows the same situation in a $\nu=1/3$ FQH bar. The hierarchy of the $\nu=1$ configuration is confirmed, with Lorentzian pulses generating minimal mixed noise for $q \in \mathbb N$ and sinusoidal voltage displaying non-optimal characteristics with non-zero $\Delta \mc S_X$. As for the charge excess noise no signature for fractional values of $q$ arises, signaling once again the robustness of levitons in interacting fractional systems. This is markedly different from driven-quantum-dot systems, where a strategy to inject a periodic train of fractionally charged quasi-particles in the FQH regime has been recently discussed \cite{Ferraro15}.

\begin{figure}
	\centering
	\includegraphics[width=\linewidth]{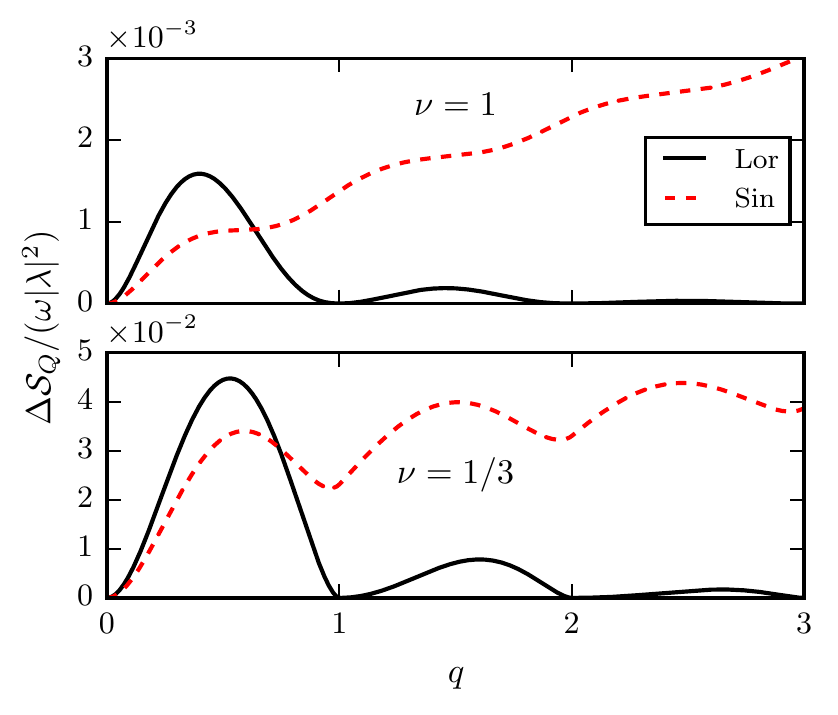}
	\caption{(Color online) Excess heat noise $\Delta \mc S_Q$ as a function of the charge per period $q$. Full black and dashed red lines represent Lorentzian and sinusoidal drives respectively. The temperature is $\theta=0$ and the cutoff is $\omega_{\rm c}=10 \, \omega$.}
	\label{fig:Sq}
\end{figure}

The same analysis can be carried out for the excess heat noise $\Delta \mc S_Q$. Equation \eqref{eq:exc_S_Q} suggests that the excess heat noise vanishes for the very same conditions that determine the vanishing of $\Delta \mc S_C$ and $\Delta \mc S_X$, given that we get a similar structure with only a different power law behavior. This expectation is confirmed in Fig.~\ref{fig:Sq}, where we report the behavior of $\Delta \mc S_Q$ for both $\nu=1$ and $\nu=1/3$. Lorentzian pulses carrying integer charge per period represent minimal-heat-noise states, independently of the filling factor.

We conclude this Section with a brief mathematical remark on the vanishing of the excess signals. Equations \eqref{eq:exc_S_C}, \eqref{eq:exc_S_X} and \eqref{eq:exc_S_Q} all share a similar structure in terms of the Fourier coefficients $p_l$, the only difference being the power law exponents $2\nu-1$, $2\nu$ and $2\nu+1$ respectively. 
Then, we can explain the common features of $\Delta \mc S_C$, $\Delta \mc S_X$ and $\Delta \mc S_Q$ by looking at the Fourier coefficient of the Lorentzian driving voltage. In such case, the analytical behavior of $e^{-i\varphi(t)}$ as a function of the complex variable $z=e^{i \omega t}$ guarantees that $p_{l<-q}=0$ when $q$ is an integer, as shown in Appendix \ref{app:fourier}. This immediately leads to the simultaneous vanishing of the three excess signals at integer charge $q$.
Let us also remark that the Lorentzian pulse is the only drive showing this striking feature, as Eqs.~\eqref{eq:exc_S_C}, \eqref{eq:exc_S_X} and \eqref{eq:exc_S_Q} all correspond to sums of positive terms and can thus \textit{only} vanish if $|p_l|^2$ is zero for all $l$ below $-q$. The only way this is possible is with quantized Lorentzian pulses.

\section{Multiple Lorentzian pulses}
\label{sec:multiple}

In the previous section we demonstrated that quantized Lorentzian pulses with integer charge $q$ represent minimal excitation states for the heat transport in the FQH regime, but this statement may potentially fail when different Lorentzian pulses have a substantial overlap. Indeed, nonlinear quantities such as $J_Q$, $\mc S_X$ and $\mc S_Q$ may behave very differently from charge current and noise, which are linear functions of the bias $V(t)$ in a Fermi liquid. For instance, at $\nu=1$ one already sees a fundamental difference between average charge and heat currents in their response to the external drive, as $J_C$ is independent of $V_{\rm AC}$, while $J_Q$ goes like $V_{\rm DC}^2 + \overline{V_{\rm AC}^2(t)}$ [see Eqs.~\eqref{eq:avg_J_C} and \eqref{eq:avg_J_Q}]. Then, one might wonder whether the independence of overlapping levitons survives when we look at such nonlinearity.
In this regard Battista {\it et al.} pointed out that in Fermi liquid systems $N$ levitons emitted in the same pulse are not truly independent excitations, since heat current and noise associated with such a drive are proportional to $N^2$ times the single-particle heat current and $N^3$ times the single-particle heat noise respectively. Nevertheless, well-separated levitons always give rise to really independent excitations with $J_Q$ and $\mc S_Q$ both equal to $N$ times their corresponding single-particle signal, due to the vanishing of their overlap \cite{Battista14}.
Moreover, an additional source of nonlinearity is provided by electron-electron interactions giving rise to the FQH phase, whose power-law behavior is governed by fractional exponents, thus strongly deviating from the linear regime.

\begin{figure}
	\centering
	\includegraphics[width=\linewidth]{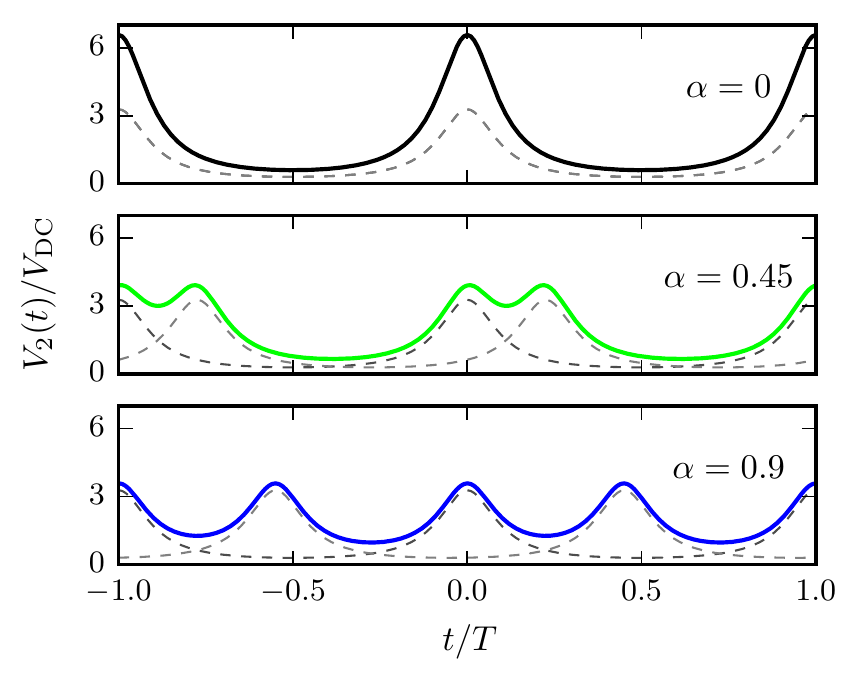}
	\caption{(Color online) Time-periodic voltage drive given by Eq.~\eqref{eq:multiple_pulses} in the case of $N=2$ Lorentzian-shaped pulses per period at total charge $q=1$ (i.e.~$1/2$ for each pulse). The top panel represents two completely overlapping pulses ($\alpha=0$), for which we simply have $V_2(t) = 2 \widetilde V(t)$. The central and bottom panels correspond to non-trivial cases $\alpha=0.45$ and $\alpha=0.9$ with finite overlap between pulses. In all cases the behavior of individual Lorentzian pulses $\widetilde V(t)$ and $\widetilde V \left(t-\frac{\alpha}{N} T\right)$ are depicted with dashed, thin lines.}
	\label{fig:multiple_pulses}
\end{figure}

In the following we study how nonlinearities due to heat transport properties and interactions affect the excess signals we introduced in Sec.~\ref{sec:excess_signal}.
For this purpose, we consider a periodic signal made of a cluster of $N$ pulses described by
\begin{equation}
	\label{eq:multiple_pulses}
	V_N(t) = \sum_{j=0}^{N-1} \widetilde V \left(t-j \frac{\alpha}{N} T\right) ,
\end{equation}
where $\widetilde V(t)$ is periodic of period $T$.
We still consider the parameter $q$ as the total charge injected during one complete period $T$ of the drive $V_N(t)$, which means that each pulse in the cluster carries a fraction $q/N$ of the total charge.
Inside a single cluster, the $N$ signals in Eq.~\eqref{eq:multiple_pulses} are equally spaced with a fixed time delay $\Delta t = \alpha T/N$ between successive pulses. Note that $\alpha=0$ corresponds to several superimposed pulses, giving $V_N(t)|_{\alpha=0}=N \widetilde V(t)$. Also, for $\alpha=1$ we just get a new periodic signal with period $T/N$. We thus restrict the parameter $\alpha$ to the interval $0 \leq \alpha < 1$. An example of such a voltage drive is provided in Fig.~\ref{fig:multiple_pulses}.

Fourier coefficients for a periodic multi-pulse cluster can be factorized in a convenient way (see Appendix \ref{app:fourier}). Here we take as an example the simple case $N=2$, whose coefficients are given by
\begin{equation}
	\label{eq:Fourier_multiple_pulses}
	p^{(2)}_l(q) = \sum_{m=-\infty}^{+\infty} e^{ i \pi \alpha m} p_{l-m} \left(\frac {q}{2} \right) p_{m} \left(\frac {q}{2} \right) ,
\end{equation}
Each pulse carries one half of the total charge $q$, a fact that is clearly reflected in the structure of Eq.~\eqref{eq:Fourier_multiple_pulses}.

\begin{figure}
	\centering
	\includegraphics[width=\linewidth]{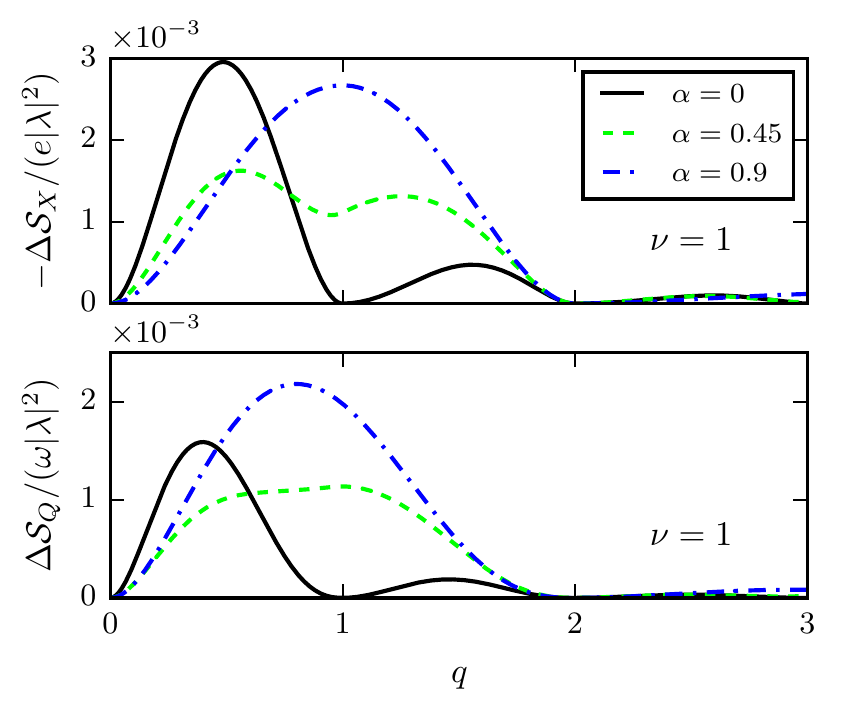}
	\caption{(Color online) Excess signals $-\Delta \mc S_X$ (top panel) and $\Delta \mc S_Q$ (bottom panel) as a function of $q$ for a cluster of two identical Lorentzian pulses separated by a time delay $\alpha T/2$. All curves refer to the case of $\nu=1$ and zero temperature. The cutoff is set to $\omega_{\rm c} = 10 \, \omega$.}
	\label{fig:Sx2_Sq2_int}
\end{figure}

Let us first focus on an integer quantum Hall effect with $\nu=1$. It is easy to see that, at least in the DC regime, $\mc S_X$ and $\mc S_Q$ scale as $V^2$ and $V^3$ respectively. It is then natural to wonder if a cluster of Lorentzian pulses still gives rise to minimal values of $\mc S_X$ and $\mc S_Q$ when the interplay of nonlinearities, AC effects and overlapping comes into play. We thus look for the excess mixed and heat noises for the case of $N=2$ Lorentzian pulses per period, in order to shed light on this problem.
The top and bottom panels of Fig.~\ref{fig:Sx2_Sq2_int} show the excess mixed and heat noises respectively in presence of two pulses per period at $\nu=1$. For $\alpha=0$ we get a perfect superposition between pulses, and we are left with a single Lorentzian carrying the total charge $q$. This case displays zeros whenever the total charge reaches an integer value, as was already discussed in the previous Section.
Higher values of $\alpha$ represent non-trivial behavior corresponding to different, time-resolved Lorentzian pulses. A Lorentzian voltage source injecting $q=1/2$ electrons per period is not an optimal drive (and so is, a fortiori, an arbitrary superposition of such pulses). As a result, signals for $\alpha=0.45$ and $\alpha=0.9$ turn out to be greater than zero at $q=1$. However $\Delta \mc S_X$ and $\Delta \mc S_Q$ still vanish at $q=2$, where they correspond to a pair of integer levitons, showing the typical behavior of minimal excitation states with no excess noise. This demonstrates that integer levitons, \textit{although overlapping}, always generate the Poissonian value for heat and mixed noises expected from their respective Schottky formulas. It is worth noticing that the blue curves in Fig.~\ref{fig:Sx2_Sq2_int} (nearly approaching the limit $\alpha \to 1$) almost totally forget the local minimum in $q=1$ and get close to a simple rescaling of the single-pulse excess noises $\Delta \mc S_X \left( \frac q 2 \right)$ and $\Delta \mc S_Q \left( \frac q 2 \right)$. This is because $\alpha \to 1$ is a trivial configuration corresponding to one pulse per period with $T'=\frac T 2$, as was mentioned before.

\begin{figure}
	\centering
	\includegraphics[width=\linewidth]{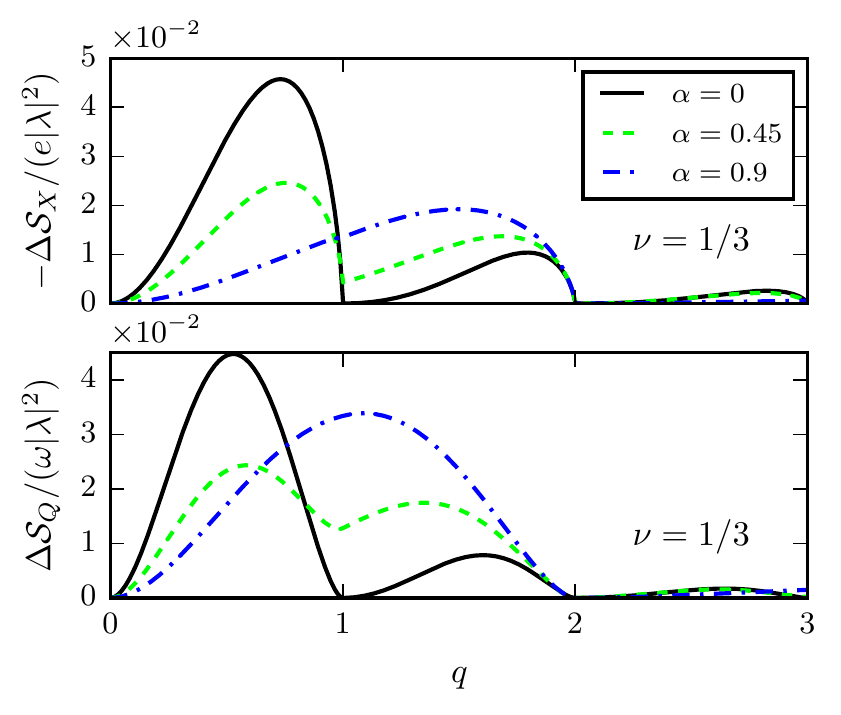}
	\caption{(Color online) Excess signals $-\Delta \mc S_X$ and $\Delta \mc S_Q$ as a function of $q$ for two identical Lorentzian pulses with time delay $\alpha T/2$ at fractional filling $\nu=1/3$ and zero temperature. The cutoff is set to $\omega_{\rm c} = 10 \, \omega$.}
	\label{fig:Sx2_Sq2_third}
\end{figure}

It is even more remarkable, however, to still observe a similar qualitative behavior in the FQH regime, where one may expect this phenomenon to break down as a result of the strong nonlinearities due to the chiral Luttinger liquid physics. Figure \ref{fig:Sx2_Sq2_third} shows that both signals drop to zero for $q=2$, representing a robust evidence for a minimal excitation state even in a strongly-interacting fractional liquid.
We stress that such a strong stability of heat transport properties is an interesting and unexpected result both at integer and fractional filling factor. Indeed, the bare signals $\overline{\lan J_Q \ran}$, $\mc S_X$ and $\mc S_Q$ are affected by the parameters governing the overlap between pulses, namely
\begin{subequations}
\label{eq:N_vs_1}
\begin{align}
	{\overline{\lan J_Q \ran}}^{(N)} & \ne N \overline{\lan J_Q \ran}^{(1)} ,\\
	{\mc S_X}^{(N)} & \ne N {\mc S_X}^{(1)} ,\\
	{\mc S_Q}^{(N)} & \ne N {\mc S_Q}^{(1)} ,
\end{align}
\end{subequations}
even at $q=N$, in accordance with Ref.~\cite{Battista14}. Nonetheless, such differences are washed out when the DC Schottky-like signals are subtracted from $\mc S_X$ and $\mc S_Q$ in Eqs.~\eqref{eq:def_exc_S_X} and \eqref{eq:def_exc_S_Q}, giving
\begin{align}
	{\Delta \mc S_X}^{(N)} & = {\Delta \mc S_X}^{(1)} = 0 ,\\
	{\Delta \mc S_Q}^{(N)} & = {\Delta \mc S_Q}^{(1)} = 0 .
\end{align}
While multiple levitons are not independent [in the sense of Eqs.~\eqref{eq:N_vs_1}], they do represent minimal excitation states even in presence of a finite overlap between Lorentzian pulses. This is a remarkable property which seems to distinguish the Lorentzian drive from every other type of voltage bias.

Let us note that the robustness with respect to the overlap of Lorentzian pulses is an interesting result for the charge transport at fractional filling as well. Indeed $\overline{\lan J_C \ran}$ and $\mc S_C$ do not show a trivial rescaling at $\nu \ne 1$. Nevertheless, we have checked that the excess charge noise $\Delta \mc S_C$ is insensitive to different overlap between levitons as it vanishes when exactly one electron is transported under each pulse, i.e.~when $q=N$. Note that a very similar behavior was described for the excess charge noise in Ref.~\cite{grenier13}, where multiple pulses were generated as a result of fractionalization due to inter-channel interactions in the integer quantum Hall regime at $\nu=2$.

\begin{figure}
	\centering
	\includegraphics[width=\linewidth]{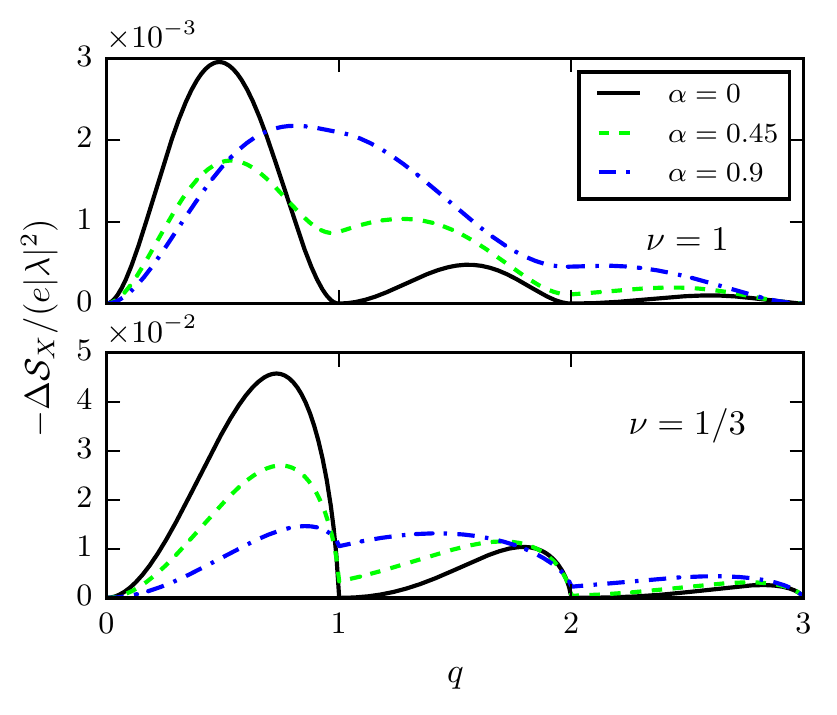}
	\caption{(Color online) Excess mixed noise $-\Delta \mc S_X$ as a function of $q$ for a cluster of two Lorentzian pulses. Here $q$ is partitioned asymmetrically, with the two pulses carrying respectively $1/3$ and $2/3$ of the total charge per period. One should compare this figure with Figs.~\ref{fig:Sx2_Sq2_int} and \ref{fig:Sx2_Sq2_third}, where $-\Delta \mc S_X$ for identical pulses is plotted. The time delay between pulses is $\alpha T/2$, with different values of $\alpha$ according to the legend. The cutoff is set to $\omega_{\rm c} = 10 \, \omega$ and the temperature is $\theta=0$.}
	\label{fig:Sx2asym}
\end{figure}

To provide a further proof for our results, we analyze a two-pulse configuration with an asymmetrical charge distribution, namely a case in which the first pulse carries $1/3$ of the total charge $q$ while the second pulse takes care of the remainder. It is straightforward to verify that the phase $e^{-i \varphi(t)}$ associated with such a drive is represented by a Fourier series with coefficients
\begin{equation}
	p^{(2)}_l(q) = \sum_{m=-\infty}^{+\infty} e^{ i \pi \alpha m} p_{l-m} \left(\frac {q}{3} \right) p_{m} \left(\frac{2q}{3} \right) ,
\end{equation}
where the asymmetry in the charge distribution is manifest, as opposed to the symmetric case in Eq.~\eqref{eq:Fourier_multiple_pulses}.
In view of previous considerations, we expect this signal to be an optimal voltage drive when both pulses carry an integer amount of charge. This condition is obviously fulfilled when $q=3$, so that the total charge can be divided into one and two electrons associated with the first and second pulse respectively. Figure \ref{fig:Sx2asym} confirms our prediction, showing the first universal vanishing point shared by all three curves at $q=3$ instead of $q=2$.

In passing, it is worth remarking that the choice of multiple Lorentzian pulses with identical shape was only carried out for the sake of simplicity. A generalization to more complicated clusters with different width $\eta$ gives rise to a very similar qualitative behavior (not shown).

\section{Conclusions}
\label{sec:conclusions}

A study of charge and heat current fluctuations and charge-heat cross-correlations in a periodically driven fractional quantum Hall system has been presented, with the goal of identifying minimal excitation states for heat transport. 
We have considered a quantum-optical protocol in which periodic voltage pulses are applied to the conductor, exciting electronic excitations. They are then scattered against a quantum point contact, leading to random partitioning of charge and heat in full analogy with the optical Hanbury Brown-Twiss experiment.
We have shown that charge, mixed and heat noises measured in one of the output arms of our interferometer all reach their minimal value (set by the respective Poissonian DC relations) when levitons impinge on the beam splitter, that is when the voltage drive generates Lorentzian pulses carrying an integer amount of electronic charge along the edge states of the quantum Hall system. These results extend the notion of leviton as a minimal excitation state in quantum conductors to the heat transport domain.
Our analysis is valid both in the integer quantum Hall effect and in the Laughlin fractional regime, despite the exotic physics due to the presence of fractionally charged quasi-particles induced by strong electron-electron interactions.

Furthermore, superposition of multiple levitons has been studied, demonstrating the robustness of levitons with respect to arbitrary overlap between them regardless of the nonlinear dependence on the voltage bias typical of heat-transport-related quantities, and despite the characteristic nonlinear power laws of the chiral Luttinger liquid theory. Our results designate levitons as universal minimal excitation states for mesoscopic quantum transport of both charge and heat.

\begin{acknowledgments}

We acknowledge fruitful discussions with D.C. Glattli. This work was granted access to the HPC resources of Aix-Marseille Universit\'e financed by the project Equip@Meso (Grant No. ANR-10-EQPX-29-01). It has been carried out in the framework of project ``1shot reloaded'' (Grant No. ANR-14-CE32-0017) and benefited from the support of the Labex ARCHIMEDE (Grant No. ANR-11-LABX-0033) and the AMIDEX project (Grant No. ANR-11-IDEX-0001- 02), all funded by the ``investissements d'avenir'' French Government program managed by the French National Research Agency (ANR).

\end{acknowledgments}

\appendix

\section{Equation of motion for the bosonic field}
\label{app:eq_motion}

In this Appendix the phase shift $e^{i e^* \int_0^{t} dt' V(t')}$ introduced in the main text is explicitly derived from the equation of motion for the field $\phi_R$. 

We consider a generic external voltage bias $\mc V(x,t)$ coupled with the right-moving density $\rho_R (x) = - \frac{\sqrt{\nu}}{2\pi} \partial_x \phi_R(x)$. Adding the capacitive coupling to the total Hamiltonian $H_0$ we get
\begin{align}
	H & = \frac{v}{4\pi} \int dx \left[ \left(\partial_x \phi_R \right)^2 + \left(\partial_x \phi_L \right)^2 \right] \nonumber \\
	& \quad - \frac{e \sqrt \nu}{2\pi} \int dx \mc V(x,t) \partial_x \phi_R(x) .
\end{align}
The equation of motion for the field $\phi_R$ is easily derived
\begin{equation}
\label{eq:eq_motion}
	(\partial_t + v \partial_x) \phi_R(x,t) = e \sqrt \nu \mc V(x,t) .
\end{equation}
Solutions of Eq.~\eqref{eq:eq_motion} are of the form
\begin{equation}
\label{eq:sol_eq_motion}
	\phi_R(x,t) = \phi_R^{(0)}(x,t) + e \sqrt \nu \int_0^t dt' \mc V[x-v(t-t'),t'] ,
\end{equation}
where $\phi_R^{(0)}(x,t)=\phi_R^{(0)}(x-vt,0)$ is the free chiral field in the equilibrium, non-biased configuration. Causality due to the propagation of excitations at finite velocity $v$ is manifest in Eq.~\eqref{eq:sol_eq_motion}.

To make contact with experiments we consider an uniformly-oscillating semi-infinite voltage contact. Indeed, in a typical experimental setup electrons travel a long way through ohmic conductors before reaching the mesoscopic system. We model this situation through the factorization $\mc V(x,t) = \Theta(-x-D) V(t)$, with $D$ the finite distance between the contact and the QPC (which is located at $x=0$) and $\Theta(x)$ the Heaviside step function. Then Eq.~\eqref{eq:sol_eq_motion} reads
\begin{equation}
	\phi_R(x,t) = \phi_R^{(0)}(x,t) + e \sqrt \nu \int_0^{t-\frac{x+D}{v}} dt' V(t') .
\end{equation}
The unimportant constant time delay $D/v$ generated by the finite distance $D$ will be neglected throughout this paper. Finally, from bosonization identity Eq.~\eqref{eq:bosonization} we get the quasi-particle field at $x=0$
\begin{equation}
	\psi_R(0,t) = \psi_R^{(0)}(0,t) e^{-i e^* \int_0^{t} dt' V(t')} ,
\end{equation}
where the phase shift $e^{i e^* \int_0^{t} dt' V(t')}$ is recovered. It is worth noticing that we dropped the label $^{(0)}$ in the main text in order to avoid a cumbersome notation.   

\section{Fourier series for the phase factor}
\label{app:fourier}

This Appendix is devoted to the Fourier analysis of the periodic signal $e^{-i\varphi(t)}$, with $\varphi(t) = e^* \int_{0}^t V_{\rm AC}(t') dt'$. Here we work with voltage drives whose DC and AC components are constrained by the request $V_{\rm min} = 0$, where $V_{\rm min}$ is the minimum value for a single pulse, but more general choices with independent DC and AC amplitudes are possible \cite{Battista14,dubois13}. We thus consider
\begin{align}
	V_{\rm sin}(t) & = V_{\rm DC} \left[1 -\cos(\omega t)\right]  ,\\
	V_{\rm Lor}(t) & = \frac{V_{\rm DC}}{\pi} \sum_{k=-\infty}^{+\infty} \frac{\eta}{\eta^2 + (t/T-k)^2} ,
\end{align}
where $\omega$ is the angular frequency and $\eta=W/T$ governs the width of each Lorentzian pulse, with $W$ the half width at half maximum.
The sinusoidal drive is used as a prototype for non-optimal voltage drives, while the Lorentzian signal fulfills the condition discussed by Levitov and collaborators for minimal excitations in quantum conductors. They both come into play through the dynamical phase $e^{-i e^* \int_0^{t} dt' V(t')} = e^{-i\varphi(t)} e^{-i \omega q t}$, with $q=e^* V_{\rm DC}/\omega$ the total charge injected during one period of the drive. The Fourier series $e^{-i\varphi(t)} = \sum_l p_l e^{-i l\omega t}$ allows to deal with the time-dependent problem as a superposition of time-independent configurations, with energy shifted by an integer amount of energy quanta $\omega$.
Coefficients for $V_{\rm sin}(t)$ are easily found to be $p_l = J_l(-q)$ \cite{crepieux04}, where $J_l$ is the Bessel functions of the first kind. For the Lorentzian case, it is convenient to switch to a complex representation in terms of the variable $z=e^{i \omega t}$. After some algebra and introducing $\gamma=e^{-2\pi \eta}$ one finds \cite{dubois13,grenier13}
\begin{equation}
	\label{eq:pl_lorentz}
	p_l = \frac{1}{2\pi i} \oint_{|z|=1} dz \, z^{l+q-1} \left( \frac{1-z\gamma}{z- \gamma} \right)^{q} .
\end{equation}
From Eq.~\eqref{eq:pl_lorentz} one can make use of complex binomial series and Cauchy's integral theorem \cite{arfken,needham} to finally get
\begin{equation}
	p_l = q \gamma^l \sum_{s=0}^\infty (-1)^{s} \frac{\Gamma(l+s+q)}{\Gamma(1+q-s)} \frac{\gamma^{2s}}{s! (s+l)!} .
\end{equation}
With the help of Eq.~\eqref{eq:pl_lorentz} one also realizes the uniqueness of the Lorentzian drive with integer $q$. Under this assumption the integrand function in Eq.~\eqref{eq:pl_lorentz} does not have any singularity outside the unit circle for $l<-q$, even at infinity. This automatically translates into $p_l=0$ for $l<-q$, hence the vanishing of the excess signals in Eqs.~\eqref{eq:exc_S_C}, \eqref{eq:exc_S_X} and \eqref{eq:exc_S_Q}.

Finally, let us briefly discuss the case of multiple pulses of Sec.~\ref{sec:multiple}. The phase accumulated for the periodic signal $V_N(t) = \sum_{j=0}^{N-1} \widetilde V \left(t-j \frac{\alpha}{N} T\right)$ is given by
\begin{align}
	\varphi_N(t)
	& = e^* \int_0^t dt' \left[ \sum_{j=0}^{N-1} \widetilde V \left(t'-j \frac{\alpha}{N} T\right) - \widetilde V_{\rm DC} \right] \nonumber \\
	& = \sum_{j=0}^{N-1} \left[\widetilde \varphi \left(t-j \frac{\alpha}{N} T\right) - \widetilde \varphi \left(-j \frac{\alpha}{N} T\right) \right] ,
\end{align}
where $\widetilde \varphi(t) = e^*\int_0^t dt' \left[ \widetilde V(t') - \widetilde V_{\rm DC} \right]$. Each phase factor $e^{-i \widetilde \varphi(t)}$ can be written as
\begin{equation}
	e^{-i \widetilde \varphi(t)} = \sum_l p_l \left(\frac q N \right) e^{-i l\omega t} ,
\end{equation}
since each pulse $\widetilde V$ involves only a fraction of the total charge $q$. The corresponding Fourier coefficients for $e^{-i \varphi_N(t)}$ read
\begin{widetext}
\begin{align}
	p^{(N)}_l(q)
	& = \exp \left[ i \sum_{j=0}^{N-1} \widetilde \varphi \left(-j \frac{\alpha}{N} T\right) \right] \int_0^T \frac{dt}{T} e^{i l \omega t} \prod_{j=0}^{N-1} e^{-i \widetilde \varphi \left(t-j \frac{\alpha}{N} T\right) } = \nonumber \\
	& = \exp \left[ i \sum_{j=0}^{N-1} \widetilde \varphi \left(-j \frac{\alpha}{N} T\right) \right]\int_0^T \frac{dt}{T} \exp (i l \omega t) \sum_{m_0=-\infty}^{+\infty} \sum_{m_1=-\infty}^{+\infty} \dots \sum_{m_{N-1}=-\infty}^{+\infty} \exp (-i m_0 \omega t) p_{m_0} \left(\frac {q}{N} \right) \nonumber \\
	& \quad \times \exp \left\{ -i m_1 \omega \left[t-\frac \alpha N T \right] \right\} p_{m_1} \left(\frac {q}{N} \right) \dots \exp \left\{ -i m_{N-1} \omega \left[t-(N-1)\frac \alpha N T \right] \right\} p_{m_{N-1}} \left(\frac {q}{N} \right) = \nonumber \\
	& = \exp \left[ i \sum_{j=0}^{N-1} \widetilde \varphi \left(-j \frac{\alpha}{N} T\right) \right] \sum_{m_1=-\infty}^{+\infty} \cdots \sum_{m_{N-1}=-\infty}^{+\infty} \exp \left\{ i \frac{2\pi}{N} \alpha \left[ m_1 + \dots + (N-1) m_{N-1} \right] \right\} \nonumber \\
	& \quad \times p_{l-m_1-\ldots-m_{N-1}} \left(\frac {q}{N} \right) p_{m_1} \left(\frac {q}{N} \right) \cdots p_{m_{N-1}} \left(\frac {q}{N} \right) .
\end{align}
\end{widetext}
As an example, coefficients for $N=2$ are given by
\begin{equation}
	p^{(2)}_l(q) = e^{i \widetilde \varphi \left( -\frac{\alpha T}{2} \right)} \sum_{m=-\infty}^{+\infty} e^{ i \pi \alpha m} p_{l-m} \left(\frac {q}{2} \right) p_{m} \left(\frac {q}{2} \right) .
\end{equation}
Note that the time-independent phase $e^{i \widetilde \varphi \left( -\frac{\alpha T}{2} \right)}$ has been omitted in Eq.~\eqref{eq:Fourier_multiple_pulses} of the main text, as it is washed out as soon as we compute the squared modulus of $p^{(2)}_l$.
Finally it is worth remarking that in the case of Lorentzian pulses with $q=N$ one should require $m_i \geq -1$ to prevent the vanishing of $p^{(N)}_l$. It follows that $p^{(N)}_l(N) = 0$ for $l<-N$ and $\Delta \mc S_C$, $\Delta \mc S_X$ and $\Delta \mc S_Q$ all vanish under these circumstances (see Figs.~\ref{fig:Sx2_Sq2_int}, \ref{fig:Sx2_Sq2_third}, and \ref{fig:Sx2asym}).

\section{Calculation of currents and noises}
\label{app:calculations}

In this Appendix we apply the Keldysh non-equilibrium contour formalism \cite{Rammer,Martin_Houches} to the calculation of currents and noises defined in Sec.~\ref{sec:current_noise}. In this framework one has
\begin{align}
	\lan J_C(t) \ran & = \frac{e^*}{2} \sum_{\eta_0} \lan T_{\rm K} \dot N_L(t^{\eta_0}) e^{-i \int_{c_{\rm K}} dt' H_{\rm t}(t')} \ran , \\
	\lan J_Q(t) \ran & = \frac 1 2 \sum_{\eta_0} \lan T_{\rm K} \dot H_L(t^{\eta_0}) e^{-i \int_{c_{\rm K}} dt' H_{\rm t}(t')} \ran \nonumber \\
	& \quad - \frac{\mu}{e^*} \lan J_C(t) \ran ,
\end{align}
with $T_{\rm K}$ the time ordering operator along the back-and-forth Keldysh contour $c_{\rm K}$, whose two branches are labeled by $\eta_0=+,-$. 
The transparency of the QPC can be finely tuned with the help of gate voltages. In the low reflectivity regime, tunneling can be treated as a perturbative correction to the perfectly transmitting setup. Then at first order in the perturbation we have
\begin{align}
	\lan J_C(t) \ran
	& = -i \frac{e^*}{2} \sum_{\eta_0,\eta_1} \int_{-\infty}^{+\infty} dt' \eta_1 \lan T_{\rm K} \dot N_L(t^{\eta_0}) H_{\rm t}({t'}^{\eta_1}) \ran \nonumber \\
	& = i e^* |\lambda|^2 \sum_{\eta_0,\eta_1} \int_{-\infty}^{+\infty} d\tau \eta_1 \sin \left[ e^* \int_{t-\tau}^t dt'' V(t'') \right] \nonumber \\
	& \quad \times \exp \left[ 2 \nu \mc G^{\eta_0 \eta_1} (\tau)\right]
\end{align}
for the charge current, with $\lambda = \Lambda/(2\pi a)$. In the last equation we explicitly showed the matrix structure of Keldysh Green's functions due to the two-fold time contour. Indeed, both $t$ and $t'$ can be placed along the forward-going or backward-going branch of $c_{\rm K}$, giving rise to the $2\times2$ matrix
\begin{equation}
	\lan T_{\rm K} \psi_{R/L}(0,t^{\eta_0}) \psi^\dag_{R/L}(0,{t'}^{\eta_1}) \ran = \frac{\exp \left[ 2\nu \mc G^{\eta_0 \eta_1}(\tau) \right]}{2\pi a}  .
\end{equation}
Similarly, the heat current reads
\begin{align}
	\lan J_Q(t) \ran
	& = -\frac{i}{2} \sum_{\eta_0,\eta_1} \int_{-\infty}^{+\infty} dt' \eta_1 \lan T_{\rm K} \dot H_L(t^{\eta_0}) H_{\rm t}({t'}^{\eta_1}) \ran \nonumber \\
	& \quad - \frac{\mu}{e^*} J_C(t)\nonumber \\
	& = i |\lambda|^2 \sum_{\eta_0,\eta_1} \int_{-\infty}^{+\infty} d\tau \eta_1 \cos \left[ e^* \int_{t-\tau}^t dt'' V(t'') \right] \nonumber \\
	& \quad \times \exp \left[\nu \mc G^{\eta_0 \eta_1} (\tau) \right] \partial_\tau \exp \left[\nu \mc G^{\eta_0 \eta_1} (\tau) \right] .
\end{align}
Green's functions along Keldysh contour are related to the real-time correlation function $\mc G(\tau) = \lan \left[ \phi_{R/L}(0,\tau) - \phi_{R/L}(0,0)\right] \phi_{R/L}(0,0) \ran$. In particular one has \cite{Rammer}
\begin{equation}
	\label{eq:matrix_G}
	\begin{pmatrix}
	\mc G^{++}(\tau) & \mc G^{+-}(\tau) \\ 
	\mc G^{-+}(\tau) & \mc G^{--}(\tau) \\
	\end{pmatrix}
	=
	\begin{pmatrix}
	\mc G(|\tau|) & \mc G(-\tau) \\ 
	\mc G(\tau) & \mc G(-|\tau|) \\
\end{pmatrix}
\end{equation}
where the bosonic correlation function $\mc G(\tau)$ at finite temperature $\theta$ reads
\begin{equation}
	\mc G(\tau) = \ln \left[ \frac{\pi \tau \theta}{\sinh (\pi \tau \theta) (1+i \omega_{\rm c} \tau)} \right]
\end{equation}
in the limit $\omega_{\rm c}/\theta \gg 1$. Using Eq.~\eqref{eq:matrix_G} we get
\begin{align}
	\lan J_C(t) \ran
	& = 2i e^* |\lambda|^2 \int_{0}^{+\infty} d\tau \sin \left[ e^* \int_{t-\tau}^t dt'' V(t'') \right] \nonumber \\
	& \quad \times \left[ e^{ 2 \nu \mc G(\tau)} - e^{ 2 \nu \mc G(-\tau)} \right] , \\
	\lan J_Q(t) \ran
	& = i |\lambda|^2 \int_{0}^{+\infty} d\tau \cos \left[ e^* \int_{t-\tau}^t dt'' V(t'') \right] \nonumber \\
	& \quad \times \left[ \partial_\tau e^{ 2 \nu \mc G(\tau)} - \partial_\tau e^{ 2 \nu \mc G(-\tau)} \right] .
\end{align}
At this stage it is useful to introduce the Fourier transform $\hat P_g (E) = \int d\tau e^{i E \tau} e^{g \mc G(\tau)}$ of the bosonic Green's function, that reads \cite{Cuniberti88,Ferraro10}
\begin{equation}
	\hat P_g(E) = \left(\frac{2\pi \theta}{\omega_{\rm c}}\right)^{g-1} \frac{e^{E/(2\theta)}}{\Gamma(g) \omega_{\rm c}} \left|\Gamma\left(\frac{g}{2}-i \frac{E}{2\pi \theta}\right)\right|^2 .
\end{equation}
Interestingly, $P_g(E)$ is nothing but the usual Fermi distribution multiplied by the effective tunneling density of states of the chiral Luttinger liquid \cite{vannucci15}. Indeed one has $\hat P_g(E) = \mc D_g(E) n_{\rm F}(-E)$, with
\begin{equation}
	\mc D_g(E) = \frac{(2\pi)^g}{\Gamma(g) \omega_{\rm c}} \left(\frac{\theta}{\omega_{\rm c}}\right)^{g-1} \frac{\left|\Gamma\left(\frac{g}{2}-i \frac{E}{2\pi \theta}\right)\right|^2}{\left|\Gamma\left(\frac{1}{2}-i \frac{E}{2\pi \theta}\right)\right|^2}
\end{equation}
and the Fermi distribution defined having zero chemical potential
\begin{equation}
	n_{\rm F}(E) = \frac{1}{1+e^{E/\theta}} .
\end{equation}
A constant tunneling density of states is recovered for the Fermi liquid case $\nu=1$, in accordance with the assumption of linear dispersion typical of the Luttinger liquid paradigm. At $\theta = 0$ we resort to the asymptotic limit of the gamma function \cite{nist} to obtain Eq.~\eqref{eq:Pg_zero_temp}
\begin{equation}
	\hat P_g(E) = \frac{2\pi}{\Gamma(g) \omega_{\rm c}} \left| \frac{E}{\omega_{\rm c}} \right|^{g-1} \Theta(E) .
\end{equation}
Using Fourier representations for $e^{ 2 \nu \mc G(\tau)}$ and $e^{-i \varphi(t)}$ we obtain
\begin{align}
	\lan J_C(t) \ran
	& = |\lambda|^2 e^* \sum_{l,m} p_l^* p_m e^{i(l-m) \omega t} \nonumber \\
	& \quad \times \left\{ \hat P_{2\nu} \left[ (q+m)\omega \right] - \hat P_{2\nu} \left[ -(q+m)\omega \right] \right\} , \\
	\lan J_Q(t) \ran
	& = \frac 1 2 |\lambda|^2 \sum_{l,m} p_l^* p_m e^{i(l-m) \omega t} (q+m) \omega \nonumber \\
	& \quad \times \left\{ \hat P_{2\nu} \left[ (q+m)\omega \right] - \hat P_{2\nu} \left[ -(q+m)\omega \right] \right\} .
\end{align}
Averaging over one period of the voltage drive we get  
\begin{align}
	\overline{\lan J_C(t) \ran}
	& = |\lambda|^2 e^* \sum_{l} |p_l|^2 \nonumber \\
	\label{eq:avg_J_C_finite_temp}
	& \quad \times \left\{ \hat P_{2\nu} \left[ (q+l)\omega \right] - \hat P_{2\nu} \left[ -(q+l)\omega \right] \right\} , \\
	\overline{\lan J_Q(t) \ran}
	& = |\lambda|^2 \frac \omega 2 \sum_{l} |p_l|^2 (q+l) \nonumber \\
	\label{eq:avg_J_Q_finite_temp}
	& \quad \times \left\{ \hat P_{2\nu} \left[ (q+l)\omega \right] - \hat P_{2\nu} \left[ -(q+l)\omega \right] \right\}
\end{align}
where the notation $\overline{\lan \dots \ran}$ stands for $\int_0^T \frac{dt}{T} \lan \dots \ran$.
Equations \eqref{eq:avg_J_C} and \eqref{eq:avg_J_Q} of the main text immediately follows when we perform the zero-temperature limit of Eqs.~\eqref{eq:avg_J_C_finite_temp} and \eqref{eq:avg_J_Q_finite_temp}.

We now turn to the calculation of the noises defined in Eqs.~\eqref{eq:def_S}. First of all, we note that all terms $\lan J_i(t) \ran \lan J_j(t) \ran$, with $i,j=C,Q$, are $O(|\lambda|^4)$, and the lowest order terms in the perturbative expansion are thus given by
\begin{multline}
	\lan T_{\rm K} \Delta J_i(t^{+}) \Delta J_j({t'}^{-}) e^{-i \int_{c_{\rm K}} d\tau H_{\rm t}(\tau)} \ran = \\
	= \lan J_i(t^{+}) J_j({t'}^{-}) \ran + O(|\lambda|^4) .
\end{multline}
Therefore, one gets the following expression for the zero-frequency charge noise
\begin{align}
	\mc S_C
	& = 2 (e^*)^2 \int_0^T \frac{dt}{T} \int_{-\infty}^{+\infty} dt' \lan T_{\rm K} \dot N_L(t^+) \dot N_L({t'}^-) \ran \nonumber \\
	& = 4 (e^*)^2 |\lambda|^2 \int_0^T \frac{dt}{T} \int_{-\infty}^{+\infty} dt' \cos \left[e^* \int_{t'}^t d t'' V(t'') \right] \nonumber \\
	& \quad \times e^{2\nu \mc G(t'-t)} ,
\end{align}
with the help of the matrix representation Eq.~\eqref{eq:matrix_G}. Mixed and heat noises are obtained in similar ways: the former reads
\begin{align}
	\mc S_X
	& = 2 e^* \int_0^T \frac{dt}{T} \int_{-\infty}^{+\infty} dt' \lan T_{\rm K} \dot N_L(t^+) \dot H_L({t'}^-) \ran  - \frac{\mu}{e^*} \mc S_C \nonumber \\
	& = 4 e^* |\lambda|^2 \int_0^T \frac{dt}{T} \int_{-\infty}^{+\infty} dt' \sin \left[e^* \int_{t'}^t d t'' V(t'') \right] \nonumber \\
	& \quad \times e^{\nu \mc G(t'-t)} \partial_{t'} e^{\nu \mc G(t'-t)} ,
\end{align}
while the latter is given by
\begin{align}
	\mc S_Q
	& = 2 \int_0^T \frac{dt}{T} \int_{-\infty}^{+\infty} dt' \lan T_{\rm K} \dot H_L(t^+) \dot H_L({t'}^-) \ran \nonumber \\
	& \quad - 2\frac{\mu}{e^*} \mc S_X + \left(\frac{\mu}{e^*}\right)^2 \mc S_C \nonumber \\
	& = 4 |\lambda|^2 \int_0^T \frac{dt}{T} \int_{-\infty}^{+\infty} dt' \cos \left[e^* \int_{t'}^t d t'' V(t'') \right] \nonumber \\
	& \quad \times e^{\nu \mc G(t'-t)} \partial_t \partial_{t'} e^{\nu \mc G(t'-t)} .
\end{align}
Using the series $e^{-i \varphi(t)} = \sum_l p_l e^{-il\omega t}$ and the Fourier transform for $e^{\nu \mc G(t'-t)}$ one is left with
\begin{align}
	\mc S_C & = 2 (e^*)^2 |\lambda|^2 \sum_{l} |p_l|^2 \nonumber \\
	& \quad \times \left\{ \hat P_{2\nu} \left[ (q+l)\omega \right] + \hat P_{2\nu} \left[ -(q+l)\omega \right] \right\} , \\
	\mc S_X & = e^* \omega |\lambda|^2 \sum_{l} |p_l|^2 (q+l) \nonumber \\
	& \quad \times \left\{ \hat P_{2\nu} \left[ (q+l)\omega \right] + \hat P_{2\nu} \left[ -(q+l)\omega \right] \right\} , \\
	\mc S_Q & = \frac{|\lambda|^2}{\pi} \sum_{l} \left|p_l\right|^2 \int_{-\infty}^{+\infty} dE E^2 \hat P_\nu(E) \nonumber \nonumber \\
	& \quad \times \left\{ \hat P_\nu \left[(q+l)\omega - E\right] + \hat P_\nu \left[-(q+l)\omega - E\right] \right\} ,
\end{align}
thus recovering Eqs.~\eqref{eq:S_C_finite_temp} and \eqref{eq:S_X_finite_temp} of the main text. To get Eq.~\eqref{eq:S_Q_finite_temp} as well, we exploit the integral \cite{nist}
\begin{align}
	& \int_{-\infty}^{+\infty} \frac{dY}{2\pi} Y^2 \hat P_{g_1}(Y) \hat P_{g_2}(X-Y) \nonumber \\
	& \quad = \frac{\hat P_{g_1+g_2}(X)}{1+g_1+g_2} \left[ g_1 g_2 \pi^2 \theta^2 + \frac{g_1 (1+g_1)}{g_1+g_2} \omega^2 \right] .
\end{align}
The latter leads to
\begin{align}
	\mc S_Q & = |\lambda|^2 \sum_{l} \left|p_l\right|^2 \left[ \frac{2\pi^2 \nu^2}{1+2\nu} \theta^2 + \frac{1+\nu}{1+2\nu} (q+l)^2 \omega^2 \right]\nonumber \\
	& \quad \times \left\{ \hat P_{2\nu} \left[ (q+l)\omega \right] + \hat P_{2\nu} \left[ -(q+l)\omega \right] \right\} .
\end{align}
One should note that for $\nu=1$ and finite temperature we have
\begin{align}
	\mc S_C & = \frac{e^2 |\Lambda|^2 \omega}{\pi v^2} \sum_{l=-\infty}^{+\infty} \left|p_l\right|^2 (q + l) \coth \frac{(q+l)\omega}{2\theta} , \\
	\mc S_X & = \frac{e|\Lambda|^2 \omega^2}{2\pi v^2} \sum_{l=-\infty}^{+\infty} \left|p_l\right|^2 (q+l)^2 \coth \frac{(q+l)\omega}{2\theta} , \\
	\mc S_Q & = \frac{|\Lambda|^2 \omega^3}{3 \pi v^2} \sum_{l=-\infty}^{+\infty} \left|p_l\right|^2 \left[\left( \frac{\pi \theta}{\omega} \right)^2 + (q+l)^2 \right] \nonumber \\
	& \quad \times (q+l) \coth \left[ \frac{(q+l)\omega}{2\theta} \right] ,
\end{align}
consistently with previous results in the literature \cite{Averin10,dubois13,grenier13,Battista14,Battista14_jpcs,moskalets14}.

\bibliography{heat_levitons_biblio}

\end{document}